\documentclass[final,3p,times,pdflatex]{elsarticle}
\usepackage{axodraw}

\bibstyle{elsarticle-num}

\def\SOFTSUSY{{\tt SOFTSUSY}}
\def\SUSPECT{{\tt SUSPECT}}
\def\SPHENO{{\tt SPHENO}}
\def\ISASUGRA{{\tt ISASUGRA}}
\newcommand{\newc}{\newcommand}
\newc{\half}{\frac{1}{2}}
\newc{\eps}{\epsilon}
\newc{\lam}{{\bf \lambda}}
\newc{\lamp}{{\bf \lambda}^{\prime}}
\newc{\lampp}{{\bf \lambda}^{\prime\prime}}
\newc{\nonr}{\nonumber}
\newc{\kap}{\kappa}
\newc{\lab}[1]{\label{eq:#1}}
\newc{\bino}{\widetilde{\cal B}}
\newc{\wino}{\widetilde{\cal W}}
\newc{\gluino}{\widetilde{\cal G}}
\newc{\code}[1]{{\tt #1}}
\newc{\ovl}{\overline}
\newc{\mlh}[1]{({m}_{\tilde{L}_{#1} H_1}^2)}
\newc{\mhl}[1]{({m}_{H_1 \tilde{L}_{#1}}^2)}
\newc{\ml}{{( m_{\tilde{L}}}^2)}
\newc{\rpv}{{\mbox{${\not\!\!R_p}$}}}

\journal{Computer Physics Communications}

\begin{document}

\begin{frontmatter}

\title{Including $R-$parity violation in the numerical computation of the
  spectrum of the minimal supersymmetric standard model: \SOFTSUSY{\tt 3.3.3}}

\author{B.C.~Allanach}
\address{DAMTP, CMS, University of Cambridge, Wilberforce road, Cambridge, CB3
  0WA, United Kingdom}

\author{M.A.~Bernhard}
\address{Physics Institute, University of Bonn, Nussallee 12, D-53115 Bonn,
Germany}
\begin{abstract}
Current publicly available computer programs calculate the
spectrum and couplings of the
  minimal supersymmetric standard model under the
  assumption of   $R-$parity conservation. Here, we describe an extension to the
  {\tt SOFTSUSY}~program which includes $R-$parity violating effects.
  The user provides a theoretical boundary
  condition upon the high-scale 
  supersymmetry breaking $R-$parity violating couplings.
  Successful radiative electroweak symmetry breaking,
   electroweak and CKM matrix data are used
  as weak-scale boundary conditions. 
  The renormalisation group equations are solved
  numerically between the weak scale and a high energy scale using a nested
  iterative algorithm. 
  This paper serves as a manual to the
  $R-$parity violating mode of the program, detailing the approximations and
  conventions used. 
\end{abstract}

\end{frontmatter}

\section{Program Summary}
\noindent{\em Program title:} \SOFTSUSY{}\\
{\em Program obtainable
  from:} {\tt http://projects.hepforge.org/softsusy/}\\
{\em Distribution format:}\/ tar.gz\\
{\em Programming language:} {\tt C++}, {\tt fortran}\\
{\em Computer:}\/ Personal computer\\
{\em Operating system:}\/ Tested on Linux 4.x\\
{\em Word size:}\/ 32 bits\\
{\em External routines:}\/ None\\
{\em Typical running time:}\/ a few seconds per parameter point.\\
{\em Nature of problem:}\/ Calculating supersymmetric particle spectrum and
mixing parameters in the $R-$parity violating minimal supersymmetric standard
model. The solution to the renormalisation group equations must be consistent
with a high-scale boundary condition on supersymmetry breaking parameters and
$\rpv$ parameters, as well as a weak-scale boundary condition on gauge
couplings, Yukawa couplings and the Higgs potential parameters.\\
{\em Solution method:}\/ Nested iterative algorithm. \\
{\em Restrictions:} {\SOFTSUSY} will provide a solution only in the
perturbative r\'{e}gime and it
assumes that all couplings of the MSSM are real
(i.e.\ $CP-$conserving). The iterative \SOFTSUSY~algorithm will not 
converge if parameters are too close to a boundary of successful electroweak
symmetry breaking, but a warning flag will alert the user to this fact.

\newpage

\section{Introduction}

Spectrum generators are a widely used tool in particle
physics beyond the Standard Model (SM), especially in the case of
Supersymmetric (SUSY) models. 
Spectrum generators can be used in theoretical studies of a SUSY
breaking scheme, for example in studies of fine-tuning. 
Phenomenological investigations of new
patterns of SUSY breaking require a calculation of the spectrum.  
Often, the resulting SUSY spectrum is used to calculate the prospects of high
energy experiments such as the Large Hadron 
Collider (LHC) discovering and measuring SUSY particles, assuming some 
SUSY breaking scheme \cite{Allanach:2008zn}. In order to run a realistic
collider 
and detector simulation of a new physics signal, a consistent model is needed
as input. Such simulations are required in order to set search and measurement
strategies~\cite{Armstrong:1994it,Bayatian:2006zz} as well as to estimate 
SUSY backgrounds to some measurement. 
In the event of discovery of some SUSY signals in LHC data, attention will
turn to the 
question of which patterns of SUSY breaking are consistent with data. 
In such tests, SUSY
spectrum generation would be an essential step.
SUSY studies often perform parameter scans, resulting in a large number of
generated spectra. 
There is  
thus a need for accurate and quick computation of the supersymmetric spectrum
as a first step in such studies. 
There exist several publicly available spectrum generators for
the $R-$parity ($R_p$) conserving minimal
supersymmetric extension of the Standard Model (MSSM):
\ISASUGRA~\cite{Paige:2003mg}, \SOFTSUSY~\cite{Allanach:2001kg}, 
\SUSPECT~\cite{Djouadi:2002ze} and
\SPHENO~\cite{Porod:2003um}. 
Spectrum information is typically
transferred to decay packages and event generators 
via a file in the SUSY Les Houches Accord
format~\cite{Skands:2003cj,Allanach:2008qq}. 

The most general renormalisable superpotential of the  MSSM
contains $R-$Parity violating ($\rpv$) 
couplings, violating
baryon and lepton number~\cite{Dreiner:1997uz}. A symmetry can be imposed upon
the model in order to maintain stability of the proton, for example baryon
triality~\cite{Ibanez:1991pr} or proton hexality~\cite{Dreiner:2007vp}.
It has been shown
that $\rpv$ models may have interesting features, such as the generation of
non-zero neutrino masses without the addition of
right-handed neutrino fields \cite{Hall:1983id}, and the gravitino as a viable
dark matter candidate \cite{Buchmuller:2007ui}. The 
violation of baryon or lepton number implied by $\rpv$ leads
to additional possibilities for SUSY detection, since such quantum numbers are 
conserved in the perturbative SM\@. There are important implications
for direct 
collider searches, since one can lose the classic large missing transverse
energy ``smoking-gun'' signature of SUSY\'. All of these features make $\rpv$
worthy of study. There is thus a strong motivation to extend the
$R_p-$conserving spectrum generating public computer programs to include
$\rpv$ effects. Here, we describe such an extension which has
been applied to \SOFTSUSY\@. The latest version of \SOFTSUSY~including
$\rpv$ effects can be downloaded from address
\begin{quote}
{\tt http://projects.hepforge.org/softsusy/}
\end{quote}
Installation instructions and more detailed technical documentation of the
code may also be found there.

The $R_p$ conserving aspects of \SOFTSUSY~are already explained in
detail in ref.~\cite{Allanach:2001kg}, and so they shall not be repeated here
or throughout this manual, which will concentrate solely on the $\rpv$ aspects of the calculation.
Adding $\rpv$ couplings roughly doubles the (already large)
number of couplings of the MSSM\@. The calculation is thus more complicated and 
so it takes considerably longer than the $R_p$ case (roughly a
factor of three for identical precision). However, this still means that a
single point in parameter space can be calculated in a couple of seconds on a
modern personal computer. 
The added complication of $\rpv$ means that some features of the 
$\rpv$ version of \SOFTSUSY~are less accurate the the $R_p-$conserving 
case, using only one-loop RGEs to evolve the couplings and masses of MSSM
fields, as opposed to two-loop RGEs in the $R_p$ case. Therefore, taking the
$\rpv$ calculation in the limit of small $\rpv$ couplings, the numerical
values of \SOFTSUSY~outputs will not {\em exactly}\/ agree with the
$R_p-$conserving 
version of \SOFTSUSY\@. 
We stress though, that if the user does not desire to include $\rpv$ couplings, 
the program automatically uses the $R_p$ calculation with the associated speed and
accuracy. Where the accuracy of \SOFTSUSY~in the $\rpv$-mode differs from the
$R_p-$conserving mode calculation, we shall make a 
note.

We proceed with a definition of the \SOFTSUSY~convention for the $\rpv$
parameters and mixings in section~\ref{sec:notation}. Next, in
section~\ref{sec:calculation}, we discuss the calculation, making a note of
parts which differ in accuracy to the \SOFTSUSY~$R_p$ calculation. 
Installation instructions can be found on the \SOFTSUSY~web-site, but
instructions to run the program can be found in Appendix~\ref{sec:run}. 
The output from a \SOFTSUSY~sample run is displayed and discussed in
Appendix~\ref{sec:output},
whereas a sample main program is shown and explained in
Appendix~\ref{sec:prog}. Some more technical information on the structure of
the program can be found in Appendix~\ref{sec:objects}. 
It is expected that the information in Appendix~\ref{sec:objects} will only be
of use to users who wish to 
`hack' \SOFTSUSY~in some fashion. 

\section{MSSM $\rpv$ Parameters \label{sec:notation}}

In this section, we introduce the $\rpv$ MSSM parameters
in the \SOFTSUSY~conventions. The translations to the actual variable
names that are being used in the program code are shown explicitly in
appendix~\ref{sec:objects}. The $\rpv$ \SOFTSUSY~calculation follows 
ref.~\cite{Allanach:2003eb} and so the notation and conventions are 
similar. 

\subsection{Supersymmetric parameters \label{susypars}}
The chiral superfield particle content of the MSSM has the 
following $SU(3)_c\times SU(2)_L\times U(1)_Y$ quantum numbers:
\begin{eqnarray}
L:&(1,2,-\half),\quad {\bar E}:&(1,1,1),\qquad\, Q:\,(3,2,\frac{1}{6}),\quad
{\bar U}:\,(\bar 3,1,-\frac{2}{3}),\nonr\\ {\bar D}:&(\bar 3,1,\frac{1}{3}),\quad
H_1:&(1,2,-\half),\quad  H_2:\,(1,2,\half).
\label{fields}
\end{eqnarray}
$L$, $Q$, $H_1$, and $H_2$ are the left handed doublet lepton and
quark superfields and the two Higgs doublets. $\bar E$, $\bar U$, and
$\bar D$ are the lepton, up-type quark and down-type quark
right-handed superfield singlets, respectively. 
Note that the lepton
doublet superfields $L^a_i$ and the Higgs doublet superfield coupling
to the down-type quarks, $H_1$, have the same SM gauge 
quantum numbers. 
The $\rpv$ part of the
renormalisable MSSM superpotential is written, in the interaction eigenbasis,
\begin{equation} 
W_{\rpv}=\eps_{ab}\left[ \frac{1}{2} \lam_{ijk} L_i^a L_j^b{\bar E}_k +
\lamp_{ijk} L_i^a Q_j^{xb} {\bar D}_{kx}  - \kap^i L_i^a H_2^b \right]+
\frac{1}{2}\eps_{xyz} \lampp_{ijk} {\bar U}_i^x{\bar
D}_j^y{\bar D}^z_k. 
 \label{superpot1} 
\end{equation} 
Here, we denote an $SU(3)$ colour index of the fundamental
representation by  $\{x,y,z\} \in \{1,2,3 \}$. The $SU(2)_L$ fundamental
representation indices are denoted by $\{a,b,c\} \in \{1,2\}$ and the generation
indices by $\{i,j,k\} \in \{1,2,3\}$. 
$\epsilon_{xyz}=\epsilon^{xyz}$ and  $\epsilon_{ab}=\epsilon^{ab}$ are totally
antisymmetric tensors, with $\epsilon_{123}=1$ and $\epsilon_{12}=1$,
respectively.  Currently, 
only real couplings in the superpotential and Lagrangian are included. 

\subsection{$\rpv$ SUSY breaking parameters \label{sec:susybreak}}

We now detail the notation of the soft $\rpv$ SUSY breaking parameters. The
trilinear $\rpv$ scalar interaction potential is
\begin{equation}
  V_{3, \rpv} = \eps_{ab} \left[\frac{1}{2}
    h_{ij k}\tilde{L}_i^a \tilde{L}_j^b
    \tilde{{e}}_k +h^\prime_{i jk} \tilde{L}_i^a \tilde{Q}_j^{bx}
    \tilde{d}_{kx} +~{\rm H.c.}
  \right]
   + \left[ \frac{1}{2}\epsilon_{xyz}h^{\prime\prime}_{ijk} \tilde{u}_i^x
  \tilde{d}_j^y \tilde{d}_k^z + {\rm H.c.} \right]
  \label{softhabertril}
\end{equation}
where fields with a tilde are the scalar components of the superfield
with the identical capital letter. The electric charges of $\tilde u$,
$\tilde d$, and $\tilde e$ are $-\frac{2}{3}$, $\frac{1}{3}$, and $1$,
respectively. ``H.c.'' denotes the Hermitian conjugate of the preceding terms.

The bilinear $\rpv$ scalar interaction potential is given by
\begin{eqnarray}
  V_{2, \rpv} &=& -\eps_{ab} D_i \tilde{L}_i^a H_2^b~+~m^2_{\tilde L_i H_1}\tilde{L}^\dagger_{ia}H_1^a +~{\rm H.c.}
  \label{softhaberbil}
\end{eqnarray}

\subsection{Tree-level masses \label{sec:tree}}

The mixing of
MSSM particles can change in the case that lepton number is violated by the
$\rpv$ interactions. Two cases of lepton number violating mixings are
implemented in \SOFTSUSY: neutrino-neutralino mixing and
chargino-lepton mixing. We neglect sneutrino-anti-sneutrino
mixing, because this has been shown to have negligible
phenomenological consequences once experimental bounds have been
applied~\cite{Dedes:2007ef}.  

In the presence of lepton number violating $\rpv$ interactions, the neutrinos
mix with the neutralinos. At tree level, this results in one massive 
neutrino, two massless neutrinos and four massive neutralinos. $\rpv-$loop
corrections to the neutral 
fermion mass matrix (currently neglected by \SOFTSUSY) can result in all
neutrinos acquiring masses and the emergence of a PMNS mixing matrix in lepton
charged current interactions. 
The ($7 \times
7$) neutrino-neutralino mass matrix for the three generations of
neutrinos is given in \cite{Allanach:2003eb} and reads  
\begin{eqnarray}
{\cal L}= -\frac{1}{2} (\nu_i, -i \bino, -i \wino^{(3)},
\tilde{h_1^0}, \tilde{h_2^0}
)
{\cal M}_{\rm N} \left ( \begin{array}{c}
\nu_j \\ -i \bino \\  -i \wino^{(3)} \\ \tilde{h_1^0} \\
\tilde{h_2^0} \end{array} \right ), 
\end{eqnarray}
where 
\begin{eqnarray}
{\cal M}_{\rm N} = \left ( \begin{array}{ccccc}
 0_{ij}& -\frac{g'}{2} v_i   &  \frac{g_2}{2} v_i   &  0   & - \kappa_i \\[4mm]
 -\frac{g'}{2} v_j    &   M_1   &   0  &  -\frac{g'}{2} v_d   &  \frac{g'}{2} v_u \\[4mm]
  \frac{g_2}{2} v_j   &   0  &   M_2   &  \frac{g_2}{2} v_d   &  -\frac{g_2}{2} v_u\\[4mm]
 0    &   -\frac{g'}{2} v_d  &  \frac{g_2}{2} v_d    &   0  & -\mu \\[4mm]
 - \kappa_i    &  \frac{g'}{2} v_u   &  -\frac{g_2}{2} v_u   & -\mu    & 0 \\[4mm]
\end{array} \right ),\label{neutralino}
\end{eqnarray}
where
$\kappa_i$ are the bilinear mixing parameters in Eq.~(\ref{superpot1}),
$v_i$ are the sneutrino vacuum expectation values (VEVs) and the $M_{1},M_2$ are
the gaugino masses of hypercharge and weak isospin, respectively. The matrix
(\ref{neutralino}) has five non-zero 
eigenvalues, i.e.\ four neutralinos and one neutrino. We denote
the mass eigenstates which are obtained upon diagonalisation of 
${\mathcal M}$ (in ascending order of mass):
$\nu_{i=1,2,3},\tilde{\chi}^0_{1,2,3,4}$, 
with masses along the diagonal of the matrix
\begin{equation}
{\mathcal M}_{\rm N}^{diag} = O^T {\mathcal M}_{\rm N} O, \label{Nmix}
\end{equation}
where $O$ is a member of $O(7)$. A simple multiplication of rows of $O$ by
factors of $i$ can absorb any minus signs in ${\mathcal M}_{\rm N}^{diag}$.

In addition,  charged leptons mix with the charginos. The Lagrangian
contains the ($5 \times 5$) lepton-chargino mass matrix
\begin{eqnarray}
{\cal L}= - ( -i \wino^-, \tilde{h}_2^-, e_{L_j}^-,
)
{\cal M}_{\rm C} \left ( \begin{array}{c}
 -i \wino^+\\ \tilde{h}_2^+ \\ e_{R_k}^+ \\
 \end{array} \right ) + {\rm H.c.}\,.
\end{eqnarray}
The mass eigenstates $\ell=(e,\mu,\tau), \tilde{\chi}^\pm_{1,2}$
are given upon the diagonalisation of the matrix
\begin{eqnarray}
{\cal M}_{\rm C} = \left ( \begin{array}{ccc}
M_2  & \frac{g_2}{\sqrt{2}}v_u & 0_j  \\
 \frac{g_2}{\sqrt{2}}v_d  & \mu & -(Y_E)_{ij}v_i\frac{1}{\sqrt{2}} \\
\frac{g_2}{\sqrt{2}}v_i  & \kappa_i & \frac{1}{\sqrt{2}}\left((Y_E)_{ij} v_d+\lambda_{kij}v_k\right)\\
 \end{array}
\right ) , \label{chargino}
\end{eqnarray}
Here, $Y_E$ is the lepton Yukawa matrix from the $R_p$
superpotential in ref.~\cite{Allanach:2001kg}. We define the diagonalised mass
matrix 
\begin{eqnarray}
{\cal M}^{diag}_{\rm C}= U {\cal M}_{\rm C} V^T, \label{diagUV}
\end{eqnarray}
$U$ and $V$ being orthogonal 5 by 5 matrices. 

\section{Calculation Algorithm \label{sec:calculation}}

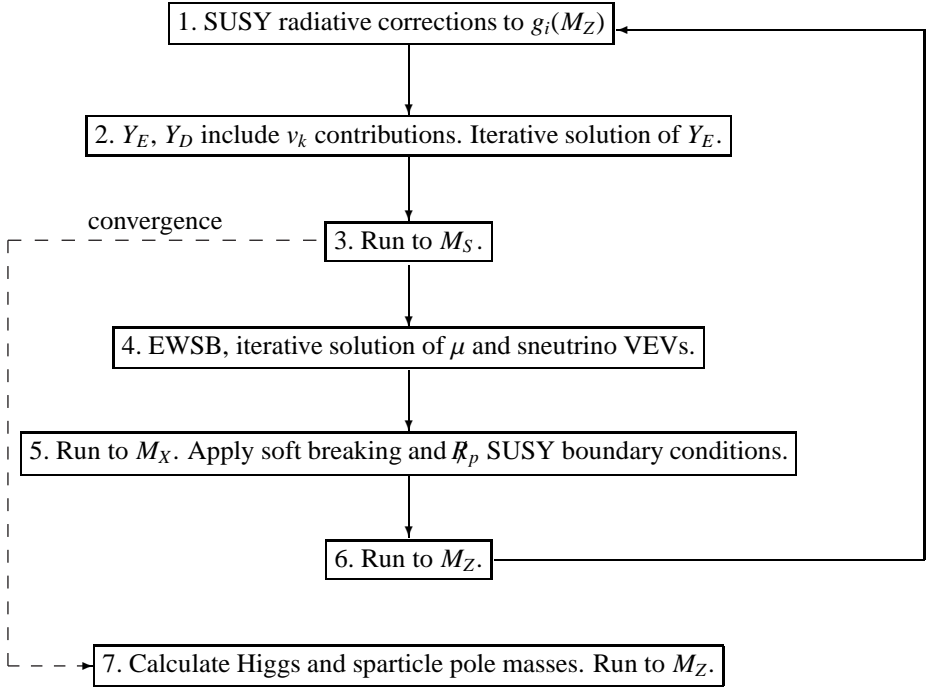
\begin{figure}
\begin{center}
\begin{picture}(323,245)
\put(10,0){\makebox(280,10)[c]{\fbox{7.\ Calculate Higgs and
      sparticle pole masses. Run to $M_Z$.}}}
\put(10,40){\makebox(280,10)[c]{\fbox{6.\ Run to $M_Z$.}}}
\put(150,76.5){\vector(0,-1){23}}
\put(10,80){\makebox(280,10)[c]{\fbox{5.\ Run to $M_X$. Apply soft breaking
and $\rpv$ SUSY boundary conditions.}}}
\put(150,116.5){\vector(0,-1){23}}
\put(10,120){\makebox(280,10)[c]{\fbox{4.\ EWSB, iterative solution of $\mu$ and sneutrino VEVs.}}}
\put(150,156){\vector(0,-1){23}}
\put(10,160){\makebox(280,10)[c]{\fbox{3.\ Run to $M_S$.}}}
\put(30,170){convergence}
\DashLine(115,165)(0,165){5}
\DashLine(0,165)(0,5){5}
\DashLine(0,5)(30,5){5}
\put(30,5){\vector(1,0){2}}
\put(150,197){\vector(0,-1){24}}
\put(150,239){\vector(0,-1){26}}
\put(60,245){\fbox{1.\ SUSY radiative corrections to
$g_i(M_Z)$}}
\put(10,200){\makebox(280,10)[c]{\fbox{2.\ $Y_E$, $Y_D$ include $v_k$
      contributions. Iterative solution of $Y_E$.}}} 
\put(182,45){\line(1,0){161}}
\put(343,45){\line(0,1){200}}
\put(343,245){\vector(-1,0){115}}
\end{picture}
\end{center}
\caption{Iterative algorithm used to calculate the $\rpv$ MSSM spectrum. 
The initial step is the
uppermost one. $M_S$ is the scale at which the EWSB
conditions 
are imposed, as discussed in the text. $M_X$ is the scale at which the high
energy SUSY breaking boundary conditions are imposed.
\label{fig:algorithm}}
\end{figure}
In broad terms, the algorithm for the calculation of the $\rpv$ MSSM spectrum
follows that of the $R_p$-conserving case,  although some of the
individual steps differ in the $\rpv$ case. It is performed via the iterative
algorithm depicted in Fig.~\ref{fig:algorithm}.
An initial estimate of
gauge couplings and up quark masses at $M_Z$ are obtained as
in the $R_p$ version of {\tt SOFTSUSY}~in ref.~\cite{Allanach:2001kg}. 
However, the charged lepton and down-quark Yukawa couplings receive $\rpv$
corrections, and are detailed in section~\ref{gyuk}. The MSSM parameters are
then run to the scale
\begin{equation}
M_S \equiv \sqrt{m_{\tilde t_1}(M_S) m_{\tilde t_2}(M_S)}, \label{msusy}
\end{equation}
where the scale dependence of the electroweak breaking conditions is
small \cite{Casas:1998cf}. Electroweak symmetry breaking (EWSB) conditions are
then imposed, as described in section~\ref{ewsb}, taking into account
sneutrino VEVs and other lepton-number violating effects. The MSSM parameters
are then run up to some high energy scale $M_X$, where the soft SUSY breaking
terms are fixed by a user-specified boundary condition. $M_X$ may be the
electroweak gauge unification scale, some scale pre-specified by the user or
indeed $M_S$. The running of the MSSM couplings is described in
section~\ref{running}. 
The SUSY $\rpv$ couplings $\lam_{ijk}$, $\lam'_{ijk}$,
$\lampp_{ijk}$, $\kappa_i$ are then fixed at this scale $M_X$.
The model is then run down to $M_Z$, when the iteration is performed again by
returning to step 1 in Fig.~\ref{fig:algorithm}. Iteration proceeds until, at
step 3, all parameters evaluated at $M_S$ are identical to within a fractional
accuracy of \code{TOLERANCE} to the previous iteration's (at step 3). 
\code{TOLERANCE}$<1$ is a numerical parameter set by the user, with default
value $10^{-3}$. 
Once this has been
achieved, the algorithm proceeds to step 7, where the pole masses of
sparticles are calculated as in section~\ref{spec}.

\subsection{Gauge and yukawa couplings \label{gyuk}}

In order to calculate the Yukawa couplings of the down quarks, contributions
to the mass matrix from sneutrino VEVs are taken into 
account:
\begin{equation}
(Y_{D})_{ij} = \frac{1}{v_d}\left[\sqrt{2}(m_D)_{ij} -\lambda'_{kij}\cdot
   {v_k}\right], 
\end{equation}
where all parameters are evaluated at $M_Z$ and are in the $\overline{DR}$
scheme in the MSSM\@. The down-quark mass matrix in the weak eigenbasis,
$(m_D)_{ij}$, is obtained as in the $R_p$ {\tt 
  SOFTSUSY}~version. $v_d$ is the VEV of $H_1$, as obtained below.

The chargino-lepton mixing in Eq.~(\ref{chargino}), complicates the
matching of $(Y_E)_{ij}$ to the charged lepton masses. We employ
an iterative procedure in order to calculate
which $(Y_E)_{ij}$ predict the empirical input values of charged lepton masses.
\begin{enumerate}
\item
Initially, we set $(Y_E)_{ij}$ as in the $R_p-$conserving limit, ignoring any
charged lepton-chargino mixing or sneutrino VEVs. Thus, the empirical MSSM
$\overline{DR}$ values of charged lepton masses evaluated at $M_Z$ are
written in the matrix $(m_E^{exp})_{ij}$ as diagonal values. Then, 
$(Y_{E})_{ll}= (m_E^{exp})_{ll} \sqrt{2} / v_d$ (no sum on $l$). 
\item
The resulting matrix $(Y_E)_{ij}$ is then to substituted  into
Eq.~(\ref{chargino}) to form
the charged
lepton-chargino mass matrix, obtaining the $5\times 5$ $U$ and $V$
transformation matrices that diagonalise it via Eq.~(\ref{diagUV}). 
\item
We denote the 3 by 3 lower right-hand side blocks of $U$ and $V$ by $\tilde U$
and $\tilde V$, respectively. The Yukawa matrix $Y_E$ is then set to be
\begin{equation}
(Y_{E})_{ij} = \frac{1}{v_d}\left[\sqrt{2}{\tilde
      U}^T_{ik}(m_E^{exp})_{kl}{\tilde V}_{li}-\lambda_{kij}\cdot v_k\right].
\end{equation}
{\em Physical}\/ lepton mixing is implemented in an extension of this
work~\cite{rpvneut}. 
\item
This result for $Y_E$ is then 
inserted back into step 2, leading to a better
approximation of $U$ and $V$. Steps 2 to 4 are iterated until successive
iterations predict identical diagonal entries of $Y_E$ within a fractional
accuracy of $10^{-4}\times $\code{TOLERANCE}.
\end{enumerate}

\subsection{Running of MSSM couplings~\label{running}}

For the $R_p-$conserving
parameters, the renormalisation group evolution (RGE) employs two-loop MSSM
$\beta$ 
functions for the supersymmetric 
parameters~\cite{Barger:1993gh}, including $\tan \beta$ and the Higgs
VEV parameter $v$. 
Gaugino masses and $R_p-$conserving SUSY breaking Higgs parameters are also
run to two-loop order in the $R_p-$conserving parameters. The other $R_p-$
conserving SUSY breaking parameters (sfermion mass matrices and some
tri-linear couplings) may be set to two-loop or one-loop order by the boolean
parameter \code{INCLUDE\_2\_LOOP\_SCALAR\_CORRECTIONS} in the main program.
The RGE includes
full family dependence and the complete set of 1-loop MSSM $\rpv$ $\beta$
functions in both SUSY-preserving and SUSY-breaking $\rpv$ 
parameters \cite{Allanach:2003eb}. 
The increased number
of $\rpv$ couplings and $\beta-$functions, as well as other complications,
mean that the $\rpv$ mode runs more slowly than the $R_p-$conserving mode. 
In the case of running \SOFTSUSY~in the $\rpv$ mode
the accuracy does not match the extremely high one of the $R_p$
version, in order to keep the running time down. 
 In either the $R_p-$conserving or the $\rpv$ mode, the program 
can be made to run faster by switching off the two-loop renormalisation of 
the scalar masses and tri-linear scalar couplings. All
$\beta$ functions are real and include inter-generational quark mixing
effects. 

\subsection{Electroweak symmetry breaking \label{ewsb}}

We now discuss the minimisation the potential of the neutral
scalar fields, $\{ h_2^0,\,h_1^0,\,{\tilde
  \nu}_1, {\tilde
  \nu}_2, {\tilde
  \nu}_3  \}$ at the renormalisation scale
$M_{S}$. Following the
calculation in 
Ref.~\cite{Allanach:2003eb}, this system of equations is solved using the
following definitions \cite{Nowakowski:1995dx} 
\begin{equation}
\tan\beta  \equiv  \frac{v_u}{v_d}, \qquad
v^2 \equiv v_u^2+v_d^2+\sum_{i=1}^3 v_i^2 = \frac{4 M_W^2}{g_2^2}.
\label{vevtotal}
\end{equation}
The VEVs $v_d$ and $v_u$ can be written 
\begin{equation}
v_d^2 =  \cos^2\beta \biggl (v^2-\sum_{i=1}^3 v_i^2 \biggr ), \qquad
v_u^2 =  \sin^2\beta \biggl (v^2-\sum_{i=1}^3 v_i^2 \biggr ).
\label{vevu} 
\end{equation}
We see from Eq.~(\ref{vevu}) that the presence of sneutrino VEVs does not change
the numerical value of 
$\tan \beta$. This convenient formulation was
first developed in Ref.~\cite{Nowakowski:1995dx}. 
The EWSB condition for the Higgs superpotential mass term $\mu$ can be written
\cite{Allanach:2003eb} as
 \begin{equation}
 |\mu|^2 \ = \frac{1}{\tan^2\beta -1} \left(
 \biggl [  \ovl{m}_{H_1}^2 + \mlh{i} \frac{v_i}{v_d} 
 +\kap_i^* \mu \frac{v_i}{v_d} \biggr ]
 - 
 \biggl [ \ovl{m}_{H_2}^2
 +|\kap_i|^2 - \frac{1}{4} (g^2+g_2^2) v_i^2 - \widetilde{D}_i
 \frac{v_i}{v_u} \biggr ] \tan^2\beta
 - \frac{1}{2}
 M_Z^2 \right), \label{muMSSM}
 \end{equation}
The soft SUSY breaking mass squared term ${m_3^2}$ is expressed in terms of
$\mu,\,v_u,\,v_d,$ and $v_i$: 
\begin{equation}
m_3^2 = \frac{\sin2\beta}{2} \biggl \{\biggl [ \ovl{m}_{H_1}^2
+\ovl{m}_{H_2}^2 + 2 |\mu|^2 +|\kap_i|^2 \biggr ]+ \biggl
[ \mlh{i}+\kap_i^*\mu \biggr ]\frac{v_i}{v_d} -\widetilde{D}_i
\frac{v_i}{v_u}
\biggr \} , \label{minV}
\end{equation}
where we employ the 
simplifying notation
\begin{equation}
\ovl{m}_{H_2}^2 \equiv  m_{H_2}^2 +\frac{1}{v_u} 
\frac{\partial {\bf \Delta V}}{\partial v_u}, \qquad
\ovl{m}_{H_1}^2 \equiv  m_{H_1}^2 +\frac{1}{v_d} 
\frac{\partial {\bf \Delta V}}{\partial v_d}.
\end{equation}
The tadpoles $\frac{\partial {\bf \Delta V}}{\partial v_{u,d}}$ currently
only contain the $R_p-$conserving contributions.  
$\rpv$ contributions to them are currently neglected, for these are of order
$\frac{\lambda^2_{ijk}}{16\pi^2}, \frac{{\lambda'}_{ijk}^2}{16\pi^2}$, so they are small
for small $\rpv$ couplings $\lambda_{ijk},\lambda'_{ijk}$ as is implied for
most $\{ i,j,k \}$ 
by the experimental bounds~\cite{Barbier:2004ez}. 
In the $R_p-$conserving limit $\kap_i,$ $v_i,\,\widetilde{D}_i,\,\mlh{i} \to
0$, Eqs.~(\ref{muMSSM}) and (\ref{minV}) tend to the usual $R_p-$conserving
MSSM Higgs 
potential minimisation conditions. 

The potential minimisation conditions for the sneutrino VEVs may be
written as
\begin{equation}
v_j = \sum_i({M_{\tilde{\nu}}^2}^{-1})_{ji}\biggl \{ -\biggl [ \mhl{i}+\mu^*\kap_i \biggr ]
v_d + \widetilde{D}_i v_u -\frac{\partial 
{\bf \Delta V}}{\partial v_i} \biggr \}, \label{minVI}
\end{equation}
where
\begin{equation}
(M_{\tilde{\nu}}^2)_{ij} \equiv \ml_{ji}+\kap_i\kap_j^* +\frac{1}{2}
M_Z^2 \cos2\beta \: \delta_{ij} 
+\frac{(g^2+g_2^2)}{4} 
\sin^2\beta\, (v^2-v_u^2-v_d^2)\: \delta_{ij}.
\end{equation}

We now detail the iterative procedure 
by which \SOFTSUSY~obtains parameters describing a
minimum of the potential with the correct properties, i.e.\ satisfying
Eqs.~(\ref{muMSSM}),(\ref{minV}) and (\ref{minVI}). All of the running
parameters discussed are evaluated at 
a renormalisation scale $Q=M_S$. 
\begin{enumerate}
\item
For given value of $\tan \beta$, 
Eq.~(\ref{vevtotal})
provides an initial estimate for $v_u$ and $v_d$ in the $R_p-$conserving limit
of $v_i=0$.  
Eqs.~(\ref{muMSSM}) and (\ref{minV}) are also first solved in the $R_p$
limit. 
\textit{i.e.} 
$v_i=0,\kap_i=\widetilde{D}_i=\mhl{i}=0$. This provides initial values
for $\mu$ and $m_3^2$. 
\item
The sneutrino VEVs $v_i$ are now obtained from the left hand side of
Eq.~(\ref{minVI})  
by using $v_u,\,v_d, \mu$ and $m_3^2$ as 
previously derived in the iterative procedure in the right hand side of the
equation. 
\item
The
corrected values of $v_u,\,v_d$ are then computed including the
non-zero sneutrino VEVs $v_i$ via Eq.~(\ref{vevu}). 
\item
$\mu$ and 
$m_3^2$ are then obtained from the left hand sides of 
Eqs.~(\ref{muMSSM}) and (\ref{minV}). The program returns to step 2 and steps
2-4 are iterated until 
$\{ v_i, \mu, m_3^2 \}$ all change by less
than a fractional accuracy of \code{TOLERANCE}$\times 10^{-4}$ on successive
iterations.  
\end{enumerate}

\subsection{MSSM spectrum \label{spec}}

Neutralino masses are calculated at tree-level as in Eq.~(\ref{neutralino}). 
All $R_p-$conserving one-loop threshold corrections are then added. The
neutrino masses are calculated by diagonalising this 
mass matrix, and taking the lightest three eigenvalues, whereas neutralino
masses are defined to be the largest four eigenvalues. 
When the chargino masses are calculated by the iterative procedure in
section~\ref{gyuk}, one-loop $R_p-$conserving corrections to
chargino masses are added to the two by two top left-hand corner of
Eq.~(\ref{chargino}). 
All $\overline{DR}$ quantities in the mass matrices are taken at the
renormalisation scale
$M_S$ in the tree-level mass matrices.  All other
masses are calculated according to the $R_p-$conserving \SOFTSUSY~calculation,
i.e.\ including the one-loop $R_p-$conserving threshold contributions.

\subsection{Physics applications}
The $\rpv$ aspects of prototype versions of \SOFTSUSY have already proved
useful for various 
studies, for example: in determining the different possibilities for the lightest
supersymmetric particle in the CMSSM
framework~\cite{Dreiner:2008ca,Bernhardt:2008jz}, in defining benchmark points
for future comparative collider studies~\cite{Allanach:2006st}, for studying
neutrino mass textures~\cite{Allanach:2007qc} and for investigating $\rpv$
mechanisms of neutrinoless double beta
decay~\cite{Allanach:2009iv,Allanach:2009xx}. 

\section*{Acknowledgments}
We thank S Grab, S Kom and P Slavich for discussions and 
comments on the manuscript.
MB thanks DAMTP, Cambridge, for repeatedly
offered, warm hospitality. This work has been partially supported by PPARC and
STFC\@. We thank S Grab, M Hannussek, S Kom, M McCullough and P Slavich for
bug-reports.

\appendix

\section{Running \SOFTSUSY}
\label{sec:run}

\SOFTSUSY~produces an executable called \code{softpoint.x}. For the calculation
of the spectrum of single points in parameter space, we recommend the
SUSY Les Houches Accord 2 (SLHA2)~\cite{Allanach:2008qq}  input/output
option. The user must provide a file (\textit{e.g.} the example file included
in the \SOFTSUSY~distribution
\code{rpvHouchesInput}), that specifies the model dependent input
parameters. The program may then be run with
\small
\begin{verbatim}
 ./softpoint.x leshouches < rpvHouchesInput
\end{verbatim}
\normalsize
For the SLHA2 input option, 
the output will also be given in 
SLHA2 format. The example file provided calculates the same point as the
CMSSM point we give as an example below. Such output can be used for
input into other programs which subscribe to the accord, such as
\code{PYTHIA}~\cite{Sjostrand:2007gs} (for
simulating sparticle production and decays at colliders), for example. For
further details on the necessary format of 
the input file, see ref.~\cite{Allanach:2008qq}. Note, that \SOFTSUSY~does not
yet support the (optional) setting of the bilinear sneutrino VEVs, 
these are instead fixed by Eq.~\ref{minVI}. It supports 
the setting of all other SLHA2 input blocks associated with non-complex $\rpv$. 
There is an option to have the boundary condition on $R-$parity violating
parameters to be set at $M_Z$, rather than at $M_{GUT}$. This is controlled by
the boolean global variable \code{susyRpvBCatMSUSY}, which if set to \code{true}
in the main program, will activate the $M_{SUSY}$ option ($M_{GUT}$ being the
default). One can instead switch the option on instead within the SLHA2 input
file by using a \SOFTSUSY~specific option in \code{Block SOFTSUSY}:
\begin{verbatim}
Block SOFTSUSY
  8   1.000000e+00 # Switch MSUSY-scale RPV boundary conditions ON
\end{verbatim}
Another option has been included in order to interface with programs that
expect output only in SLHA 1 format, rather than SLHA 2 format. For this,
another \code{Block SOFTSUSY}~option
\begin{verbatim}
 10   1.000000e+00 # Try to output object in SLHA 1 format 
\end{verbatim}
will attempt to produce $\rpv$ output close to the SLHA1 format. 

For a quick examination of a single point in CMSSM parameter space, the
command  
\small
\begin{verbatim}
./softpoint.x sugra <m0> <m12> <a0> <tanb> <mx> <sgnMu> RPVcoupling <i> <j> <k> <value>
\end{verbatim}
\normalsize
can be utilised. Bracketed entries should be replaced by their numerical
values, where all massive parameters ($m_0$, $M_{1/2}$, $A_0$, $M_{GUT}$)
should be 
quoted in GeV. \code{RPVcoupling}~$\in \{$\code{lambda}, \code{lambdaP}, or
\code{lambdaPP}$\}\equiv \{\lam$, $\lam'$, $\lam''\}$ and \{\code{i,j,k}\}$\in
\{1,2,3\}$ specify a single non-zero $\rpv$ GUT-scale coupling. 
\code{mx}~denotes the scale at which the high-energy boundary condition is to
be applied. 
If this is 
specified as \code{unified}, as in the $R_p$ version, the electroweak gauge
unification scale
$M_{GUT}$ is used (defined to be the $\overline{DR}$ scale $Q$ at which $g_1(Q) =
g_2(Q)$). The default output is in SLHA2 format, the conventions of which are
explained in Ref.~\cite{Allanach:2008qq}.

\section{Sample Program \label{sec:prog}}

In this section we present a sample main program, that illustrates 
a scan over an $\rpv$ parameter. This main program
is included in the  
\code{rpvmain.cpp}~file with the standard \SOFTSUSY~distribution and performs
a scan in $\lambda'_{331}(M_{GUT})$, assuming the CMSSM10.1.1~\cite{bench}
CMSSM parameters $m_0=125$ GeV, $M_{1/2}=500$ GeV, $A_0=0$, $\tan
\beta=10$ and $\mu>0$. The  
size of the coupling varies from $\lambda''_{323} \in [0 , 0.7 ]$.
The program prints out the value of the right-handed stop pole mass and any
problems 
associated with the point in question for each value of
$\lambda''_{323}(M_{GUT})$. 

The sample program has the following form: 
\small
\begin{verbatim}
#include <rpvmain.h>

int main() {
  /// Sets up exception handling
  signal(SIGFPE, FPE_ExceptionHandler); 

  bool gaugeUnification = true, ewsbBCscale = false;

  /// Do we include 2-loop RGEs of *all* scalar masses and A-terms, or only the
  /// scalar mass Higgs parameters? (Other quantities all 2-loop anyway): the
  /// default in SOFTSUSY 3 is to include all 2-loop terms, except for RPV,
  /// which is already slow and calculated to less accuracy than the R-parity
  /// conserving version
  bool INCLUDE_2_LOOP_SCALAR_CORRECTIONS = false;

  /// Sets format of output: 6 decimal places
  outputCharacteristics(6);

  /// Header  
  cerr << "SOFTSUSY" << SOFTSUSY_VERSION 
       << " Ben Allanach, Markus Bernhardt 2009\n";
  cerr << "If you use SOFTSUSY, please refer to B.C. Allanach, ";
  cerr << "Comput. Phys. Commun. 143 (2002) 305, hep-ph/0104145;\n";
  cerr << "For RPV aspects, B.C. Allanach and M.A. Bernhardt, ";
  cerr << "Comp. Phys. Commun. 181 (2010) 232, ";
  cerr << "arXiv:0903.1805.\n";

  /// "try" catches errors in main program and prints them out
  try {

    /// Contains default quark and lepton masses and gauge coupling
    /// information 
    QedQcd oneset;      ///< See "lowe.h" for default parameter definitions 
    oneset.toMz();      ///< Runs SM fermion masses to MZ
    
    /// Print out the Standard Model data being used, as well as quark mixing
    /// assumption and the numerical accuracy of the solution
    cerr << "Low energy data in SOFTSUSY: MIXING=" << MIXING << " TOLERANCE=" 
         << TOLERANCE << endl << oneset << endl;
    /// set parameters
    double tanb = 10.;
    int sgnMu = 1;
    double mgutGuess = 2.e16; 
    double a0 = 0.0, m12 = 500.0, m0 = 125.0; 
    
    /// number of points for scan
    const int numPoints = 20; 
    
    /// parameter region
    double Start = 0. , End = 0.7;
    
    DoubleVector pars(3);
    /// set basic entries in pars
    pars(1) = m0; pars(2) = m12; pars(3) = a0;
      
    cout << "l''_{323}(M_X) m_stop_R     # Problem flag" << endl;
    /// loop over parameter space region
    int ii; for (ii=0; ii<=numPoints; ii++){
      double lambda = Start + ((End - Start) / double(numPoints) * double(ii));
      
      /// define rpvSoftsusy object
      RpvSoftsusy kw;
      
      /// set lambda coupling at mgut
      kw.setLamPrimePrime(3, 2, 3, lambda); 
      
      /// output parameters into double vector pars used by lowOrg
      kw.rpvDisplay(pars);
      
      /// generate spectrum in RpvSoftsusy object kw
      kw.lowOrg(rpvSugraBcs, mgutGuess, pars, sgnMu,
       tanb, oneset, gaugeUnification, ewsbBCscale);
      
      /// outputs for this scan
      int pos;
      cout << lambda << "  " << kw.displayPhys().mu(2, 3) << " # " 
           << kw.displayProblem() << endl;
    }
  }
  catch(const string & a) {
    cout << a; exit(-1);
  }
  catch(const char *a) {
    printf("%s", a); exit(-1);
  }
  
}
\end{verbatim}
\normalsize
After including a header file, global variables are defined. These are all
described in the $R_p$ manual~\cite{Allanach:2001kg}.
After setting the output accuracy, 
the program output begins with a title print-out. Then follow
some variables specifying the Standard Model input parameters, the
\code{MIXING}~switch, which determines how any quark mixing is implemented
(as in ref.~\cite{Allanach:2001kg}) and
the iteration precision of the output, \code{TOLERANCE}.
The running masses of the SM fermions and the QED and QCD gauge couplings are
determined at $M_Z$ from data with the method \code{toMz}.
If the switch \code{gaugeUnification=true}, \SOFTSUSY~will
determine \code{mGutGuess}~as the scale $M_{GUT}$ of electroweak gauge
unification. In order to do 
this, it requires an initial guess, which must be 
supplied as the initial value of the variable \code{mGutGuess} (in GeV), 
and is later over-written by the program\footnote{If the user wishes to provide
  this, $2 \times 10^{16}$ GeV is a good initial
guess for $M_{GUT}$.}.

The next step is the definition of the CMSSM parameters $A_0/$GeV$=$\code{a0},
$M_{1/2}/$GeV$=$\code{m12}, $\tan \beta=$\code{tanb} and
$m_0/$GeV$=$\code{m0}. Next, a \code{for}~loop performs the scan over
$\lambda''_{323}(M_{GUT})$. 
In the example given (CMSSM), the first three parameters are
\code{pars(1) = m0; pars(2) = m12; pars(3) = a0;}. The \code{pars} vector is
needed to keep track 
of the boundary conditions set at $M_{GUT}$.
In the iterative \SOFTSUSY~algorithm
the parameters in the \code{RpvSoftsusy} 
object change due to
the RGE running\@.
they are re-set in
every iteration at $M_{GUT}$ from the unchanged \code{DoubleVector pars}~
parameters. Users should note that for lepton number violating couplings,
users should use the updated \code{RpvNeutrino}~object as in
Ref.~\cite{rpvneut}, rather than a 
\code{RpvSoftsusy}~one.

We do not fill the other 102 $\rpv$ entries
of \code{pars} explicitly. This would be tedious and an additional
source of potential bugs. Instead, we fill the \code{RpvSoftsusy} object
itself using the \code{setLamPrimePrime} method in this example.  
We use the \code{rpvDisplay} method: this fills the
\code{pars} vector automatically with what was set already inside the
\code{RpvSoftsusy} object, while leaving the first nine entries in the
vector unchanged. The \code{rpvDisplay} method automatically changes the length
of \code{pars} appropriately. 
After this, the actual \SOFTSUSY~main driving method \code{lowOrg} is
called, the first argument specifying the type of boundary condition 
(currently \code{rpvSugraBcs}), which assumes that \code{pars}~has already
been prepared by using the \code{rpvDisplay} object. 
This is followed by the output of the GUT-scale coupling $\lambda$, 
the pole right-handed stop mass and any problems in the calculation of the
parameter point.  Finally, the \code{catch}~commands print any errors produced
by the program.

\section{Sample Output \label{sec:output}}

We now present some non-SLHA2 compliant \SOFTSUSY~output
for the example program presented in Section~\ref{sec:prog}.
\normalsize
The output obtained from this command is:
\small
\begin{verbatim}
# l''_{323}(M_X) m_stop_R     # Problem flag
0.000000e+00  8.147269e+02 # 
3.500000e-02  8.120697e+02 # 
7.000000e-02  8.041222e+02 # 
1.050000e-01  7.916904e+02 # 
1.400000e-01  7.740020e+02 # 
1.750000e-01  7.559414e+02 # 
2.100000e-01  7.368196e+02 # 
2.450000e-01  7.177035e+02 # 
2.800000e-01  6.993199e+02 # 
3.150000e-01  6.821923e+02 # 
3.500000e-01  6.665102e+02 # 
3.850000e-01  6.525457e+02 # 
4.200000e-01  6.399977e+02 # 
4.550000e-01  6.289228e+02 # 
4.900000e-01  6.191508e+02 # 
5.250000e-01  6.105389e+02 # 
5.600000e-01  6.029625e+02 # 
5.950000e-01  4.809702e+02 # [ Quasi-fixed point breached Non-perturbative ]
6.300000e-01  4.740464e+02 # [ Quasi-fixed point breached Non-perturbative ]
6.650000e-01  4.683686e+02 # [ Quasi-fixed point breached Non-perturbative ]
7.000000e-01  4.636366e+02 # [ Quasi-fixed point breached Non-perturbative ]
\end{verbatim}
\normalsize
After a header line labelling the contents of the columns, we see the GUT
scale value of $\lambda''_{323}$ assumed, then the pole value of the
right-handed stop mass and any problems associated with the parameter point
being examined. For large values of $\lambda''_{323}$, a quasi-fixed point
occurs in the renormalisation group equations, and no perturbative solution to
the RGEs exists. 

\section{Object Structure\label{sec:objects}}

We now go on to sketch the objects and their relationship to each other. This
is necessary information for any generalisation beyond the $\rpv$ MSSM\@. Only
methods and data which are deemed of possible importance for prospective users
are 
mentioned here, but there are many others within the program itself.

\subsection{Tensor}
\label{tensor}

The \SOFTSUSY~program comes with its own linear algebra classes of vectors and
matrices (e.g.\ \code{DoubleMatrix}) that 
have been introduced in \cite{Allanach:2001kg}.
New in this version is the class \code{Tensor}, given in files
\code{tensor.h} and \code{tensor.cpp}. This class has been added to
implement the three-index tensors containing some of the $\rpv$ couplings
into the program. For this reason, the class is specifically designed as a
vector of three single objects, each of type \code{DoubleMatrix}. The
dimension of \code{Tensor} is $(3,3,3)$. The class also
contains linear algebra functions for multiplication, addition or subtraction
with matrices and vectors via over loaded operators. For 
more detail we refer the interested reader to 
the technical documentation on the \SOFTSUSY~web-site.
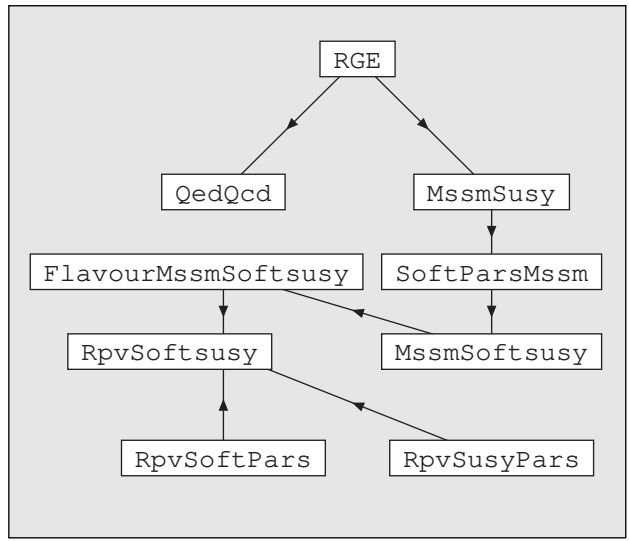
\begin{figure}
\begin{center}
\begin{picture}(215,200)(-30,0)
\GBox(200,200)(-30,0){0.9}
\ArrowLine(100,180)(50,130)
\ArrowLine(100,180)(150,130)
\ArrowLine(150,130)(150,100)
\ArrowLine(150,100)(150,70)
\ArrowLine(150,70)(50,100)
\ArrowLine(50,100)(50,70)
\ArrowLine(50,30)(50,70)
\ArrowLine(150,30)(50,70)
\SetPFont{Teletype}{10}
\BText(100,180){RGE}
\BText(50,130){QedQcd}
\BText(150,130){MssmSusy}
\BText(50,30){RpvSoftPars}
\BText(40,100){FlavourMssmSoftsusy}
\BText(30,70){RpvSoftsusy}
\BText(150,100){SoftParsMssm}
\BText(150,70){MssmSoftsusy}
\BText(150,30){RpvSusyPars}
\end{picture}
\end{center}
\caption{Heuristic high-level object structure of \SOFTSUSY\@. Inheritance is
displayed by the direction of the arrows. \label{fig:objstruc}}
\end{figure}
The \SOFTSUSY~internal representation of the trilinear $\rpv$ couplings
is different to the user interface. While the user interface uses the
common $\lambda_{ijk} L_i L_j \bar E \in W_{\rpv}$ notation, this coupling is
internally represented in the \cite{Allanach:2003eb} \code{Tensor}~notation in
terms of 
three matrices as in Eq.~\ref{trilins}. 
\begin{equation}
(\code{Lambda\_U}_i)_{jk}\equiv \lambda''_{ijk}, \qquad
(\code{Lambda\_D}_i)_{jk} \equiv \lambda'_{jki}, \qquad
(\code{Lambda\_E}_i)_{jk} \equiv \lambda_{jki}. \label{trilins}
\end{equation}
This does not stop
the user from only operating 
the program using the user interface and the usual $\lambda_{ijk}$ conventions.
$\lambda'_{jki}$, $\lambda''_{ijk}$ are also stored within analogous
\code{Tensor}~representations of three 3$\times$3 matrices. 

\subsection{General object structure}

From an RGE point of view, data in a particular quantum field theory 
consist of a set of parameters defined at some
renormalisation scale $Q$. 
A set of $\beta$ functions describes the
evolution of the parameters and masses to a different scale
$Q'$. This concept is embodied in an {\em abstract} \code{RGE}
object, which contains the methods required to run objects of derived
classes to different renormalisation scales (their beta functions). The other
objects 
displayed in figure~\ref{fig:objstruc} are particular instances of
\code{RGE}, and therefore inherit from it. \code{QedQcd}, \code{MssmSusy},
\code{SoftParsMssm} and \code{MssmSoftsusy}~objects encode the $R_p$ part of the
MSSM and its SM input data~\cite{Allanach:2001kg}.
\code{RpvSusyPars} contains all of the supersymmetric $\rpv$ couplings contained
within  Eq.~(\ref{superpot1}). 
\code{RpvSoftPars}~contains the $\rpv$ soft supersymmetry breaking parameters
listed in 
Eqs.~(\ref{softhabertril}) and~(\ref{softhaberbil}). \code{RpvSoftsusy} is
the $\rpv$  
generalisation of the \code{MssmSoftsusy}~class, and contains all $\rpv$ MSSM
couplings along with their beta functions.
Methods for the \code{RpvSoftsusy}~class exist to perform the minimisation of
the neutral scalar potential as well as the
calculation of Yukawa couplings described in section~\ref{sec:calculation}. 
Neutrino, neutralino and chargino masses and mixings are calculated within
this class. 
Code in
the \code{MssmSoftsusy} class organises and performs the main part of
the calculation, using polymorphism to detect the correct $\beta$ functions to
use (in this case, the $\rpv-$MSSM $\beta$ functions). 
All of the \code{Rpv}~objects contain default constructors
and destructors, as well as overloaded \code{>>},\code{<<}~operators for input
and output. 
There is always an implicit dependence of running RGE quantities on the
current renormalisation scale $Q$. Thus, if a method is called that returns
one of the object's couplings or masses, that object will return it at the
current scale $Q$ of the object. 
In the following, we provide basic information on the 
classes associated with $\rpv$, so that users may program using the class
structure of \SOFTSUSY\@. More detailed and
technical documentation on the program should be obtained from the
\SOFTSUSY~website.

\subsection{\code{RpvSusyPars}~class}

\begin{table}\begin{center}
\begin{tabular}{lll}
  data variable & & methods \\ \hline
  \code{\small Tensor lu, ld, le} & trilinear $\rpv$ superpotential &
  \code{\small displayLam}, \code{\small displayLamPrime}
 \\
  $\Lambda_E, \Lambda_D, \Lambda_U$ & couplings & \code{\small displayLamPrimePrime}
  \\ 
 & & \code{\small setLam}, \code{\small setLamPrime} \\ 
  & & \code{\small setLamPrimePrime}\\\hline
  \code{\small DoubleVector kappa} & bilinear $\rpv$ parameter&
  \code{\small displayKappa}\\  
  $\kappa_{i=1,2,3}/$GeV &  & \code{\small setKappa} \\
\hline \normalsize
\end{tabular}\end{center}
\caption{\code{RpvSusyPars} class data and accessor methods. See
  Eq.~(\protect\ref{trilins}) for a translation between the
  structure of   
  \code{Tensor lu, ld, le}. \label{tab:rpvsusy}}
\end{table}
Each of the higher level
objects described in this appendix have explicitly named
\code{display}  and \code{set} methods that are used to access or
change the data contained within each object. In
table~\ref{tab:rpvsusy} (as in the following tables in this
section), these accessing methods are listed on the same row as the
relevant data variable. The data and input/output methods in the \code{RpvSusyPars} class are
presented in table~\ref{tab:rpvsusy}.
When using \code{Tensor}~objects
\code{lu,ld,le}, there exists an enumerated type
  \code{RpvCouplings} $\in$ $\{$\code{LU, LD, 
  LE}$\}$~used as arguments to the \code{displayLambda,setLambda} methods.
This argument selects the type of coupling ($\lambda''_{ijk}$,
$\lambda'_{ijk}$ or $\lambda_{ijk}$). 

\subsection{\code{RpvSoftPars}~class}

\begin{table}\begin{center}
\begin{tabular}{lll}
  data variable & & methods \\ \hline
    \code{\small DoubleVector mH1lsq} & bi-linear scalar 
  & \code{\small displayMh1lSquared} \\ 
$m^2_{\tilde L_{i=1,2,3} H_1}/$GeV$^2$    & parameters & \code{\small setMh1lSquared} \\ \hline
  \code{Tensor her, hdr, hur } & trilinear $\rpv$ scalar &  \code{\small displayHr} \\
  $\{h_{ijk}, h'_{ijk}, h''_{ijk}\}/$GeV &interactions & \code{\small setHr}\\
$i,j,k \in \{1, 2, 3 \}$ & & \\
\hline
 \code{\small DoubleVector dr} & bilinear $\rpv$ scalar & \code{\small
   displayDr} 
  \\ 
$D_{i=1,2,3}/$GeV$^2$    &parameters   & \code{\small setDr} \\ \hline
\end{tabular}\end{center}
\caption{\code{RpvSoftPars} class data and accessor
  methods. All parameters are running parameters, evaluated at the
  $\overline{DR}$ scale $\mu$.\label{tab:RpvSoftPars}}
\end{table}
The data and input/output methods in the \code{RpvSoftPars} class are
presented in table~\ref{tab:RpvSoftPars}. The \code{displayHr},
  \code{setHr}~methods take a parameter of the enumerated type
  \code{RpvCouplings}~as their   first 
  argument  to select a
  particular tri-linear scalar interaction that is ($h''_{ijk}$,
  $h'_{ijk}$   or $h_{ijk}$ depending upon the argument). 

\subsection{\code{RpvSoftsusy}~class}

\begin{table}\begin{center}
\begin{tabular}{lll}
  data variable & & methods \\ \hline
  \code{\small DoubleVector snuVevs} & sneutrino VEVs&
  \code{\small displaySneutrinoVevs} \\ 
$v_{i=1,2,3}/$GeV & & \code{\small setSneutrinoVevs} \\ \hline
  \code{\small DoubleVector nuMasses} & neutrino masses & \code{\small setNeutrinoMasses} \\
$m_{\nu_i={1,2,3}}/$GeV  &  &  \code{\small displayNeutrinoMasses}
\\ \hline 
  \code{\small DoubleMatrix neutralFermMixing} & neutral fermion & \code{\small displayNeutralMixing} \\
$O$ (7 by 7) &mixing  & \code{\small setNeutralMixing} \\  \hline
  \code{\small DoubleMatrix Uch} & charged fermion & \code{\small displayUch}
  \\
$U$ (5 by 5) &mixing  & \code{\small setUch} \\ \hline
  \code{\small DoubleMatrix Vch} & charged fermion & \code{\small displayVch}
  \\
$V$ (5 by 5) &mixing  & \code{\small setUch} \\ \hline
 & & \\
  method & \multicolumn{2}{l}{function} \\ \hline
  \code{\small void rpvDisplay (DoubleVector)} & \multicolumn{2}{l}{returns all 102
  $\rpv$ MSSM running parameters as a }\\
  &  \multicolumn{2}{l}{\code{\small DoubleVector}, leaving the first 9 entries untouched}
  \\ 
\hline
  \code{\small void rpvSet (DoubleVector)} & \multicolumn{2}{l}{sets all 102 $\rpv$ MSSM
  parameters from a user-provided }\\
  &  \multicolumn{2}{l}{\code{\small DoubleVector}, starting from its 10$^{th}$ entry} \\ 
\hline
\end{tabular}\end{center}
\caption{\code{\small RpvSoftsusy} class. The parameters $\nu_i, m_{{\nu_i}},
  O,U,V$   are
  tree-level $\overline{DR}$ parameters. \label{tab:RpvSoftsusy}}
\end{table}
The data and important methods in the \code{RpvSoftsusy} class are
presented in table~\ref{tab:RpvSoftsusy}. 
As well as standard constructors and destructors, there exists a
  method \code{beta}, that calculates the numerical values of the
  $\beta-$function\@. The \code{rpvDisplay}~method is used in the example program
  \code{rpvmain.cpp}, and fills a vector with MSSM running parameters in a
  certain   order. 
 \code{rpvSet}~is used to set MSSM running parameters according to a
 \code{DoubleVector}~argument, assuming the same order as \code{rpvDisplay}.
 If the user
  wishes to provide their own function encoding high-scale boundary conditions
  on the soft supersymmetry breaking terms, they must provide a function
\begin{verbatim}
void (*boundaryCondition)(MssmSoftsusy &, const DoubleVector &)
\end{verbatim}
which is then passed as the first argument to the main driving method 
\code{lowOrg}~\cite{Allanach:2001kg}. 
One can specify CMSSM conditions for
the $R_p$ parts, plus all specified $\rpv$ interactions at the high scale 
by using \code{rpvSugraBcs}. Alternatively, one can specify
$R_p-$conserving gauge mediated supersymmetry breaking conditions plus $\rpv$
interactions at the scale $M_{mess}$ by using \code{rpvGmsbBcs}. In this case,
the first elements of \code{pars}~should contain $n$, the number of vector
like 5-plets of 
messenger fields, the messenger mass scale $M_{mess}$ in GeV, $\Lambda$ and
$C_{grav}$, the constant that determines the gravitino
mass~\cite{Ambrosanio:1997rv}.  \code{rpvAmsbBcs}~implements the
minimal anomaly mediated supersymmetry breaking~\cite{Randall:1998uk} 
assumption, but {\em neglects} $\rpv$ couplings in the high-scale boundary
condition. For this case, the first two parameters of \code{pars} should be
\code{m32} and \code{m0}, respectively.
This approximation ought to be reasonable for small dimensionless
$\rpv$ couplings.
Users should note that for lepton number violating couplings,
users should use the updated \code{RpvNeutrino}~object as in
Ref.~\cite{rpvneut}, rather than a 
\code{RpvSoftsusy}~one.

\end{document}